\documentclass[fleqn,12pt,twoside]{article}
\usepackage[headings]{espcrc1}

\readRCS
$Id: espcrc1.tex,v 1.2 2004/02/24 11:22:11 spepping Exp $
\usepackage{graphicx}

\def\bfp{{\bf p}}

\usepackage[figuresright]{rotating}

\newcommand{\AmS}{{\protect\the\textfont2
  A\kern-.1667em\lower.5ex\hbox{M}\kern-.125emS}}

\title{Particle spectra and hydro-inspired models}

\author{Wojciech Florkowski \thanks{This work was supported in part by 
the Polish State Committee of Scientific Research, Grant No. 2 P03B 05925.} 
\\ \medskip
Institute of Physics, \'Swi\c{e}tokrzyska Academy \\
ul. \'Swi\c{e}tokrzyska 15, 25-406 Kielce, Poland  \\ \medskip
and \\ \medskip
H. Niewodnicza\'nski Institute of Nuclear Physics, Polish Academy of Sciences \\
ul. Radzikowskiego 152, 31-543  Krak\'ow, Poland 
}

\runauthor{W. Florkowski}

\begin{document}

\maketitle

\begin{abstract}
Several popular parameterizations of the freeze-out conditions in ultra-relativistic 
heavy-ion collisions are shortly reviewed. The common features of the models, 
responsible for the successful description of hadronic observables, are outlined.
\end{abstract}

\section{INTRODUCTION}

In this talk I present several models which turned out to be quite successful 
in reproducing the experimentally measured hadron spectra in nucleus-nucleus 
collisions. The discussed models use concepts borrowed from relativistic 
hydrodynamics but they do not include the complete time evolution of the
system. For this reason, they may be called hydro-inspired (freeze-out) 
models. The measured particle spectra reflect properties of matter at the stage 
when particles stop to interact. This moment is called the kinetic (thermal)
freeze-out. Hydro-inspired models help us to verify the idea that matter, just 
before the kinetic freeze-out, is locally thermalized and exhibits collective 
behavior, such as the transverse and longitudinal expansion.
If this is really the case,  i.e., if the hydro-inspired models describe the data
well, we may infer the thermodynamic properties of matter at freeze-out, 
such as the values of the temperature and flow, and request that the advanced 
hydrodynamic models reproduce this configuration. 

In my opinion, the real aim of the freeze-out models is to form a simple 
link between sophisticated hydrodynamic calculations \cite{Teaney:2001av,%
Huovinen:2002fp,Kolb:2002ve,Kolb:2003dz,Hirano:2004er,Socolowski:2004hw}
describing the full time evolution of matter (with the inclusion of 
the phase transition) and the rich bulk of the experimental data describing soft phenomena. This is an appealing idea, however, one encounters several problems on the way to achieve this task, since certain features of the hydrodynamic models and the freeze-out models are quite different. For example, at the first sight one can realize that typical shapes of the freeze-out hypersurfaces used in the advanced hydro calculations and in the freeze-out models are quite different.
Clearly, further work is necessary to make the connection between the 
hydro calculations and the freeze-out models more consistent.

In any case, an attractive feature of the hydro-inspired models is that they are 
very effective parameterizations of the final state, that use few parameters 
possessing clear physical interpretation. As long as we do not all have 
hydrodynamic codes that we are able to run on our PCs, the freeze-out models form 
a very convenient and easy accessible tool to interpret the data. In this talk I 
intend to review several models which give consistent description 
of many observables (not only of the particle spectra). Their list includes:
different versions of the blast-wave model \cite{Schnedermann:1993ws,Retiere:2003kf}, 
the Buda-Lund model \cite{Csorgo:1995bi,Csorgo:1999sj,Csanad:2003qa},
the Seattle model \cite{Cramer:2004ih,Miller:2005ji},
the Durham model \cite{Renk:2004yv,Renk:2004cj}, 
the Cracow single-freeze-out model \cite{Broniowski:2001we,Broniowski:2001uk,%
Broniowski:2002nf}, and THERMINATOR \cite{Kisiel:2005hn}.

Below, I will refer only to the Au+Au collisions studied at the top RHIC 
energies; I want to show how different models describe the same set of data
rather than to show how one model is able to describe different physical
situations. Examples of the application of the thermal approach to
calculate the spectra at lower energies may be found in Refs. 
\cite{Torrieri:2000xi,Letessier:2000jh}.

\section{COOPER-FRYE FORMULA AND EMISSION FUNCTION}

Before going into the discussion of the blast-wave model, I would like to make 
a few remarks about the Cooper-Frye formula \cite{Cooper:1974mv}. It is
frequently used in the hydrodynamic calculations to obtain the momentum distribution 
of the emitted particles. In this talk, the Cooper-Frye formula will serve us as  
a reference point for the characteristics of different freeze-out models. 
Its standard form is 
\begin{equation}
E_p {dN \over d^3p} =  \int \,p^\mu  d\Sigma_\mu(x) 
\, f_{\rm eq}\left(p \cdot u(x)\right),
\label{CF}
\end{equation}
where $E_p = \sqrt{m^2+{\bf p}^2}$ is the energy of a particle, $d\Sigma^\mu(x)$ a 
three-dimensional element of the freeze-out hypersurface, $u^\mu(x)$ is the 
hydrodynamic flow, and $f_{\rm eq}$ is the equilibrium distribution function. 
In the general case, the hypersurface includes the time-like (defined here by the condition: $d\Sigma^\mu d\Sigma_\mu = (d\Sigma_0)^2 - (d\Sigma_1)^2 - (d\Sigma_2)^2  - (d\Sigma_3)^2   > 0$) and space-like ($ d\Sigma^\mu d\Sigma_\mu  < 0$) parts.
The advanced hydrodynamic calculations include both parts, while hydro-inspired
models include typically only the time-like parts. The success of the hydro-inspired
models may indicate that the contributions from the space-like parts are negligible.
On the other hand, detailed microscopic studies show that the contributions from 
those parts are substantial and difficult to include in the consistent way, 
unless one uses the approach based on the kinetic theory 
\cite{Bugaev,Anderlik:1998et,Magas:1999zm,Sinyukov:2002if,Bugaev:2002ch,Magas:2003yp,%
Bugaev:2004kq,Grassi:2004dz,Molnar:2005gy,Molnar:2005gx,Tamosiunas:2005pe}. 
\begin{figure}
\includegraphics[width=15.0cm]{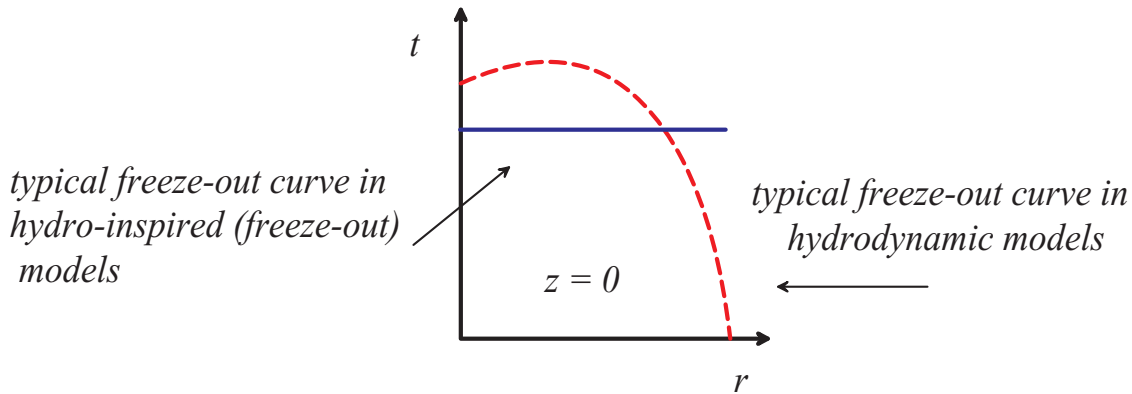}
\caption{Freeze-out curves in the Minkowski space, i.e., projections of the 
freeze-out hypersurfaces on the plane $z=0$ for cylindrically symmetric systems, in
the advanced hydro calculations (dashed line) and in the freeze-out models 
(solid line).}
\label{fig:fcurve}
\end{figure}

The Cooper-Frye formula may be rewritten in such a way that the distribution
of the particles in the momentum space is given as the space-time integral
over the so-called emission or source function
\begin{eqnarray}
E_p {dN \over d^3p} =   \int d^4x 
\int \,p^\mu  d\Sigma_\mu(x^\prime) \delta^4(x^\prime-x)
\, f_{\rm eq}\left(p \cdot u(x^\prime)\right)   
 \equiv  \int d^4x \, S(x,p).
\label{EF}
\end{eqnarray}
The emission function defines the space-time distribution of the points from
which the observed hadrons are emitted. Very often Eq. (\ref{EF}) is used with 
$S(x,p)$ modeled without any reference to the Cooper-Frye formula. Such a procedure 
lays emphasis on the fact that the hydro-inspired models should aim mainly at the 
reconstruction of the realistic emission function.

\section{BLAST-WAVE MODELS}

Different versions of the blast-wave model originate from the paper 
by Siemens and Rasmussen \cite{Siemens:1978pb}, where a relativistic formula 
for the particle distribution corresponding to a thermalized and radially
expanding system was first given. More recent applications use the same concepts
but different geometries of the expansion, more suitable for the
description of the ultra-relativistic heavy-ion collisions, are considered 
\cite{Florkowski:2004tn}.

\subsection{Cylindrically symmetric systems}

For boost-invariant and cylindrically symmetric systems, the Cooper-Frye 
formula (\ref{CF}) leads to the very popular model of Schnedermann,
Sollfrank and Heinz \cite{Schnedermann:1993ws}. For constant transverse flow, 
$v_T$ = tanh$\rho$ = const, one obtains the rapidity and transverse-momentum
distribution of the emitted particles in the form
\begin{eqnarray}
{dN \over dy d^2p_T} &=&
{e^{\beta \mu} \over  2 \pi^2} m_T
K_1 \left[\beta m_T \hbox{cosh}(\rho) \right]  
I_0 \left[\beta p_T \hbox{sinh}(\rho) \right]
\int\limits_0^1 d\zeta \,\, r(\zeta) t(\zeta) 
{dr \over d\zeta} 
\nonumber \\
&&  - {e^{\beta \mu} \over  2 \pi^2} p_T
K_0\left[\beta m_T \hbox{cosh}(\rho) \right] 
I_1 \left[\beta p_T \hbox{sinh}(\rho) \right]
\int\limits_0^1 d\zeta \,\, r(\zeta) t(\zeta) 
{dt \over d\zeta},
\label{KandI}
\end{eqnarray}
where $m_T = \sqrt{m^2+p_T^2}$ is the transverse mass, $\beta = 1/T$ is the 
inverse temperature, $\mu$ is the chemical potential, $\rho$ is the transverse 
rapidity, $t$ and $r =\sqrt{r_x^2+r_y^2}$  are the coordinates of the freeze-out 
hypersurface at $z=0$, while $K_{0,1}$ and $I_{0,1}$ are the modified Bessel 
functions. The integrals on the right-hand-side of Eq. (\ref{KandI}) involve
the parameterizations of the freeze-out position coordinates and times
in terms of the parameter $\zeta$. In practical applications, the second line on 
the right-hand-side of Eq. (\ref{KandI}) is very often neglected,
which means that the variations of the emission times are ignored.
This procedure implicitly denotes that the particle emission takes place
at a constant laboratory time (at $z=0$), see Fig.~\ref{fig:fcurve}.

An example of the procedure outlined above is the 
blast-wave fit performed recently by the BRAHMS Collaboration
\cite{Arsene:2005mr}. The optimal value of the temperature
obtained from the fit to the Au+Au data ($\sqrt{s_{NN}}$ = 200 GeV, 
midrapidity region, centrality class 0-10\%) is 110 MeV, 
while the average transverse flow velocity equals 0.65~$c$. In this, and other
similar cases, we have to be aware that a theoretical boost-invariant
model has been applied to the system which is not boost-invariant as a whole.
In my opinion, such procedure is justified only at the top RHIC energies 
in the midrapidity region where the rapidity distribution may be 
regarded as flat within 1 or at most 2 units of rapidity.

\subsection{Cylindrically non-symmetric systems}

For boost-invariant and cylindrically non-symmetric systems one may use
the parameterization of the emission function introduced by Reti\`{e}re and Lisa
\cite{Retiere:2003kf}. It takes into account the possible ellipsoidal shape 
of the system created in non central collisions.
The emission function proposed in \cite{Retiere:2003kf} may be obtained from
the Cooper-Frye formula if the constant evolution time,
$\tau = \tau_0$ = const, is replaced by a gaussian distribution 
\begin{equation}
S(x,p) = Z \, m_T\cosh(\eta-y) \, \Omega(r,\phi_s) 
\, e^{\frac{-(\tau-\tau_0)^2}{2\Delta\tau^2}} 
\, \frac{1}{e^{p\cdot u/T} \pm 1}. 
\end{equation}
Here $Z$ is an arbitrary normalization constant, $\eta = 1/2 \ln \, (t+z)/(t-z)$ 
is the spacetime rapidity, $\tan \phi_s = r_y/r_x$, $\Delta\tau$ is the emission time, 
and $\Omega$ describes the spatial distribution of matter in the transverse plane
\begin{equation}
\Omega(r,\phi_s) = \Omega(\tilde{r}) = \frac{1}{1+e^{(\tilde{r}-1)/a_s}},  
\hspace{1.0cm}
\tilde{r}(r,\phi_s) \equiv \sqrt{\frac{(r\cos(\phi_s))^2}{R_x^2}
+\frac{(r\sin(\phi_s))^2}{R_y^2}}.
\end{equation}
The parameter $a_s$ describes the surface diffuseness of the source.
In the limit $a_s \to 0$, the matter in the transverse plane is confined in an 
ellipse defined by the parameters $R_x$ and $R_y$. The explicit form of the flow
field $u^\mu$ (and of the expression $p \cdot u$) is given in Ref. 
\cite{Retiere:2003kf}, here we only note that the  expansion velocity in the 
transverse plane is perpendicular to the elliptical shell confining the matter.

The optimal values of the temperature obtained by Reti\`{e}re and Lisa in 
their analysis of three different centrality classes (0-5\%, 15-30\%, 60-92\%,
the Au+Au collisions at the beam energy of $\sqrt{s_{NN}}$ = 130 GeV, combined
PHENIX and STAR data \cite{Adcox:2001mf,Adler:2002uv}) are: 107, 106, and 100 MeV, 
respectively. The corresponding values of the transverse flow are: 0.52~$c$, 
0.50~$c$, and 0.47~$c$.

\subsection{Summary on the blast-wave models}

When applied to the Au+Au collisions at the top RHIC energies, the blast-wave models 
give good description of the $p_T$-spectra, the elliptic flow, and the HBT 
radii at midrapidity. An interesting feature of the Reti\`{e}re-Lisa model is
that it includes, as the special case, the situation where the in-plane flow is not stronger than the out-of-plane flow, but the correct sign and magnitude of the coefficient $v_2$ is reproduced. This feature is realized by the geometry of the model which implies in this case that more matter flows in the reaction plane than out of the reaction plane. 

The blast-wave fits to the 
data indicate short evolution times, $\tau \leq 10$ fm, and very short emission 
times, $\Delta \tau \leq 1$ fm. The typical value of the temperature 
is about 100 MeV.  The blast-wave models do not predict the absolute 
normalization of the spectra, hence an extra parameter is required to normalize 
each considered spectrum. In addition, many of the applications of the blast-wave 
model do not include the effects of the decays of resonances. 

\section{BUDA-LUND MODEL}

The standard version of the Buda-Lund model was formulated to describe
cylindrically symmetric systems with no constraint of the boost-invariance
\cite{Csorgo:1995bi,Csorgo:1999sj}. Later the model was extended to
describe ellipsoidally symmetric systems \cite{Csanad:2003qa}. The standard
emission function of the model has the form
\begin{equation}
S(x,p) =  {g \over (2 \pi)^3} \,
\, {  m_T \cosh(\eta - y) \over
\exp\left({u^{\mu}(x) p_{\mu} \over  T(x) } -
{\mu(x) \over  T(x)} \right) \pm 1} \, \frac{1}{(2 \pi \Delta\tau^2)^{1/2}}
\exp\left[-\frac{(\tau - \tau_0)^2} {2  \Delta \tau^2} \right],
\end{equation}
where the temperature and chemical potential depend on the position coordinates
\begin{equation}
{\mu(x) \over T(x)} =  
{\mu_0 \over T_0} - {r^2 \over 2 R_G^2}
-{\eta^2 \over 2 \Delta \eta^2 }, \label{BL1}
\end{equation}
\begin{equation}
{1 \over T(x)}  =  {1 \over T_0 } \,\, 
\left( 1 + {T_0 - T_s \over T_s} \, {r^2 \over 2 R_G^2} \right) \,
\left( 1 + { T_0 - T_{\rm e} \over T_{\rm e} }\, 
{(\tau - \tau_0)^2 \over 2 \Delta\tau^2  } \right). \label{BL2}
\end{equation}
In Eq. (\ref{BL2}), the quantity $T_0$ is the temperature in the center at the mean freeze-out time, $T_{\rm s}$ is the temperature on the surface at the mean freeze-out time, and  $T_{\rm e}$ is the temperature in the center at the end of particle emission. The flow pattern assumed in the Buda-Lund model has a Hubble-like structure, with the velocity of the fluid element proportional to its distance from the center. Such patterns appear in the analytic \cite{Csorgo:2003rt} and numerical
\cite{Chojnacki:2004ec} solutions of the equations of the relativistic
hydrodynamics. For physical interpretation of other model parameters we refer 
to the original papers \cite{Csorgo:1995bi,Csorgo:1999sj}.

The Buda-Lund model gives a very good description of the $p_T$-spectra, $v_2$ and HBT radii in the full rapidity range. The fits to the experimental data indicate small values of the evolution and emission times which are compatible with those obtained from the blast-wave analysis (for Au+Au at $\sqrt{s_{NN}}$ = 200 GeV the optimal Buda-Lund fit yields: $T_0$ = 196 MeV, $T_{\rm e}$ = 117 MeV, $T_s$ = 89.7 MeV, $R_G$ = 13.5 fm, $R_s$ = 12.4 fm, $\tau_0 =$ 5.8 fm, $\Delta \tau =$ 0.9  fm, and $\Delta \eta$ = 3.1). The effects of the decays of resonances are not included explicitly and the information about the normalization of the spectra is derived from the HBT data. The Buda-Lund fits indicate a freeze-out temperature distribution, with an average freeze-out temperature of 100 - 120 MeV. However, about 1/8th of the particles is found to be emitted from a very hot center with $T \sim$ 200 MeV. A comparison of this result with the lattice QCD results for the critical temperature was considered as an indication of quark deconfinement in Ref. \cite{Csanad:2004mm}.

\section{SEATTLE MODEL}

The Seattle model formulated in Refs. \cite{Cramer:2004ih,Miller:2005ji} describes 
cylindrically symmetric systems. No assumption about the boost-invariance is made. 
The main
characteristic feature of the model is inclusion of the effects of the final-state 
interaction of the outgoing pions. Such effects are taken into account by the use
of the distorted wave functions $\Psi_{\bfp}^{(-)}(x)$. The formalism, following 
Ref. \cite{Gyulassy:1979yi}, is based on the generalized emission function, which 
may be used to get the one-particle and two-particle distributions in the momentum 
space 
\begin{equation}
S(x,p,q) =  \int d^4K' S_0(x,K')
\int {d^4x'\over(2\pi)^4}\; e^{-i K'\cdot x'} \Psi_{\bfp_1}^{(-)}(x+x'/2) 
\Psi_{\bfp_2}^{(-)*}(x-x'/2),
\end{equation}
\begin{equation}
p=(p_1+p_2)/2, \quad  q=p_1-p_2.
\end{equation}
The quantities $p_1$ and $p_2$ are the pion four-momenta. In the special case, where 
the distorted wave functions are replaced by the plane waves, the standard 
formulation is recovered with the emission function reduced to $S_0$,
\begin{equation}
\Psi_{\bfp}^{(-)}(x) \to e^{i p \cdot x}, \quad  
S(x,p,q) \to S_0(x,p) e^{i q\cdot x}, \quad  S(x,p,0) \to S_0(x,p). 
\end{equation}
The function $S_0$ resembles the parameterization used in the Buda-Lund model,
however, in this case the thermodynamic parameters are constant,
\begin{equation}
S_0(x,p)= \frac{ m_T \cosh\eta}{\sqrt{2\pi(\Delta \tau)^2}}
            \exp \left[- {(\tau-\tau_0)^2 \over 2 \Delta \tau^2}
           -{\eta^2 \over 2 \Delta\eta^2}\right]
 {\Omega(r) \over (2\pi)^{3}}
{1\over \exp({p \cdot u- \mu_\pi \over T })-1  }.
\label{ST1}
\end{equation}
The distribution of matter is characterized by the function 
\begin{equation}
\Omega(r)= {1 \over \left[\exp\left({r-R_{WS} \over a_{WS} }\right)
+1\right]^2}, \quad r = \sqrt{r_x^2 + r_y^2}.
\end{equation}
The optimal values of the parameters obtained from the fit to the STAR data
\cite{Adams:2003xp} describing the Au+Au collisions at $\sqrt{s_{NN}} = 200$ GeV
are the following \cite{Miller:2005ji}: $T$ = 215 MeV, $\mu_\pi$ = 123 MeV,
$\tau_{0}$ = 8 fm, $\Delta\tau$ = 2.7 fm, $R_{WS}$ = 12 fm, $a_{WS}$  = 0.8 fm,
$\Delta \eta$ = 1, $\eta_f$ = 1.5, $w_0$ = 0.142$\pm$0.046 fm$^{-2}$, and 
$w_2$ = 0.582 $\pm$ 0.014~+~$i$(0.123$\pm$0.002).
The parameter $\eta_f$ defines the magnitude
of the transverse flow \cite{Cramer:2004ih}, whereas
the parameters $w_0$ and $w_2$ define the optical potential which modifies
the outgoing pion wave functions. The strength of the attraction inside the 
medium is greater than $m^2_\pi$, hence a pion behaves to large extent 
as a massless particle. This behavior may be related to the phenomena
discussed in Refs. \cite{Cohen:1994nq,Son:2001ff}.

The Seattle model describes successfully the pion transverse-momentum spectra
and the HBT radii in the region $p_T > 100$ MeV. On the other hand, 
the model predicts non-monotonic structures in the transverse-momentum 
region below 70 MeV. Such predictions were confronted with the PHOBOS data in Ref.
\cite{Back:2005vd} and certain discrepancies between the data and the model were 
found. Very likely the non-monotonic structures are due to the sudden change
of the optical potential in the transverse direction and a better agreement
with the data may be achieved by the modification of this dependence.

\section{DURHAM MODEL}

The Durham model, proposed by Renk in Refs. \cite{Renk:2004yv,Renk:2004cj},  
is based on the parameterization of the hydrodynamic expansion in the proper-time 
interval $\tau_0 \leq \tau \leq \tau_f$. The pressure profiles determine the 
evolution of the boundaries of the  system, then the boundaries 
determine the evolution of the entropy density. In the next step, the equation 
of state is used to calculate the pressure. The iterative procedure ensures that
this approach is self-consistent, i.e., the pressure profiles which determine 
the evolution of the system agree with the pressure obtained from the equation 
of state.  The freeze-out happens at a fixed temperature $T_f$ and the Cooper-Frye 
formula is used to calculate the spectra. The optimal parameters obtained from the 
fit to the PHENIX data  \cite{Adler:2003cb,Adler:2004rq} include: 
$\tau_0$ = 1~fm, $\tau_f$ = 19 fm, $T_f$ = 110 MeV. The Durham model describes 
nicely the $p_T$-spectra and HBT radii. No resonance decays are included in the 
calculation of the spectra. In my opinion, a direct comparison with the real 
hydrodynamic calculations will strengthen the importance of this model, if
the validity of the proposed parameterization is confirmed.

\section{CRACOW SINGLE-FREEZE-OUT MODEL}

The standard version of this model \cite{Broniowski:2001we} 
describes boost-invariant and cylindrically symmetric systems. The model is
based on the extreme assumption that the chemical and kinetic freeze-outs
coincide. The thermodynamic parameters of the model are obtained from
the analysis of the ratios of hadron abundances \cite{Florkowski:2001fp}. 
In this respect the Cracow model is closely related to different statistical models 
developed recently by many groups \cite{Braun-Munzinger:2001ip,Rafelski:2004dp,%
Yen:1998pa,Gazdzicki:1998vd,Cleymans:2004pp,Becattini:2003wp}. An important feature 
of the model is that a complete set of hadronic resonances is included in the 
calculation of various physical observables. The resonances are included by the 
use of the emission function that corresponds to the Cooper-Frye formula convoluted 
with the momentum  splitting functions $B$ and the space-time displacement 
functions $\delta$, 
\begin{eqnarray}
&& S\left( x_{1},p_{1}\right)  =  E_1 {dN_1 \over d^3p_1 d^4x_1 } =
\label{npix1p1} \nonumber \\ 
&& \int \frac{d^{3}p_{2}}{E_{p_{2}}}
B\left( p_{2},p_{1}\right) \int d\tau _{2}\Gamma _{2}e^{-\Gamma _{2}\tau
_{2}} \int d^{4}x_{2}\delta ^{\left( 4\right) }
\left( x_{2}+\frac{p_{2}\tau_{2}}{m_{2}}-x_{1}\right)...  \nonumber \\
&&   \hspace{5mm} \times \int d\Sigma _{\mu }\left(
x_{N}\right) \,p_{N}^{\mu }\,\,\delta ^{\left( 4\right) }\left( x_{N}+\frac{%
p_{N}\,\tau _{N}}{m_{N}}-x_{N-1}\right) f_{N}\left[ p_{N}\cdot u\left(x_{N}\right) 
\right]
\end{eqnarray}
Here the indices $1,...,N$ denote particles in a cascade of decaying
resonances ($N$ refers to the resonance decaying on the freeze-out hypersurface,
while $1$ refers to the final observed particle, e.g., a pion, other symbols are
defined in \cite{Broniowski:2001we,Broniowski:2001uk,Broniowski:2002nf}).
The freeze-out hypersurface is defined by the conditions: 
\begin{equation}
\tau_{\rm inv} = \sqrt{t^2-r^2_x-r^2_y-r^2_z} = {\rm const}, \quad 
r=\sqrt{r_x^2+r_y^2} < r_{\rm max} 
\end{equation}
\begin{equation}
u^\mu = \frac{x^\mu}{\tau_{\rm inv}} = \frac{t}{\tau_{\rm inv} } 
\left(1, 
\frac{r_x}{t}, \frac{r_y}{t}, \frac{r_z}{t}\right), \quad 
d\Sigma^\mu = u^\mu\, \tau_{\rm inv}^{3} \, {\rm sinh}(\rho)
{\rm cosh}(\rho) \, d\rho
d\eta d\phi.
\end{equation}
The standard version of the model uses only 4 parameters: temperature ($T$),
baryon chemical potential ($\mu_B$), $\tau_{\rm inv}$, and $r_{\rm max}$. 
The values of the temperature and baryon chemical potential, obtained from the 
analysis of the ratios of hadron abundances, are independent of centrality. 
For the top RHIC energies one finds $T = 165.5$ MeV and $\mu_B = 28.5$ MeV. 
The expansion parameters $\tau_{\rm inv}$ and $r_{\rm max}$ are determined
by the fits to the $p_T$-spectra, and their values depend on centrality
\cite{Baran:2003nm}.

The Cracow model gives a good description of the ratios of hadronic observables, 
$p_T$-spectra, the HBT radii $R_{\rm side}$ and $R_{\rm out}$. It also gives
satisfactory description of the balance functions \cite{Bozek:2003qi}
and of the invariant masses of the pion pairs \cite{Broniowski:2003ax}. 
Generalizations of the model describing systems without cylindrical symmetry 
\cite{Broniowski:2002wp} or boost-invariance \cite{Florkowski:2004tn} were 
proposed to account for the observed values of the coefficient $v_2$ and 
the $p_T$-spectra at finite values of the rapidity as 
measured by BRAHMS \cite{Bearden:2004yx}.

\section{THERMINATOR, SHARE, THERMUS}

THERMINATOR \cite{Kisiel:2005hn} is the Monte-Carlo version of the Cracow and 
blast-wave models (the blast-wave model has been extended in this case 
to include the complete set of hadronic resonances). The program uses
the same input from the Particle Data Tables \cite{Hagiwara:2002fs} as SHARE 
\cite{Torrieri:2004zz}. With the physical input of the thermodynamic
and expansion parameters, THERMINATOR delivers the full space-time
information about cascades of decaying resonances. The values of the thermodynamic 
parameters may be taken from SHARE or from other programs used to study the 
ratios of hadronic abundances, for example from THERMUS \cite{Wheaton:2004qb}.
The information about the space-time positions of the produced hadrons may
be used to study different types of correlations. Moreover, the Monte-Carlo 
approach facilitates the inclusion of the experimental cuts and acceptance. 

\section{HUMANIC'S RESCATTERING MODEL}

In the end of this talk I would like to note that simple transport models may 
play a similar role as hydro-inspired models. An example of such approach is
Humanic's model \cite{Humanic:2002iw}. This model is a
Monte-Carlo simulation of the evolution of purely hadronic system. The 
initialization stage may be considered as equivalent to the hadronization process.
The transverse geometry is determined by the overlapping region of the two 
colliding nuclei, while the initial $p_T$- and $y$-distributions are assumed to 
have the following shape
\begin{equation}
{1\over m_T} {dN \over dm_T} = C {m_T \over \exp(m_T/T) \pm 1}, \quad
{dN \over dy} = D e^{-{(y-y_0)^2 \over 2 \sigma_y^2}},
\end{equation}
with the optimal parameters: $T$ = 300 MeV and $\sigma_y$ = 2.4. The longitudinal 
positions and times of the initialized hadrons are: 
$z_{\rm had} = \tau_{\rm had} \,\hbox{sinh} y$ and
$t_{\rm had} = \tau_{\rm had} \,\hbox{cosh} y$, where $\tau_{\rm had}$ 
is the initial (proper) time, with the standard value of 1 fm.  

The subsequent evolution of the system includes binary collisions of:
$\pi$, $K$, $N$, $\Delta$, $\Lambda$, $\rho$, $\omega$, $\eta$,
$\eta^\prime$, $\phi$ and $K^*$.  
The freeze-out happens (dynamically) at about 30 fm (Au+Au, 130 GeV).
The model gives successful description of the slope parameters, $v_2(m,p_T,\eta)$, 
and the $p_T$ and centrality dependence of the HBT radii. 
The results are sensitive to $\tau_{\rm had}$;
larger values of $\tau_{\rm had}$ imply fewer collisions and the
rescattering-generated flow is reduced.

\section{CONCLUSIONS}

A common feature of the hydro-inspired models is the short
evolution time, $\tau_0 < $ 10~fm, and even shorter emission time
$\Delta \tau <<  \tau_0$. Another common feature is a large value of the
transverse flow, $\langle v_t \rangle \sim 0.5$ c. Altogether, these 
observations indicate an explosive scenario for the Au+Au collisions observed 
at the top RHIC energies.

Models which do not include resonances or pion final-state interactions,
yield rather low values of the freeze-out temperature, $T \sim 100$ MeV.
We know, however, that the effects of the resonance decays must be important,
since large abundances of certain resonances were observed 
\cite{Fachini:2004jx,Markert:2005ms}. 
In this situation, I think that the hydro-inspired models should enter
a new stage where the effects of both the transverse flow and the resonance 
decays are commonly taken into account. Another development of the
hydro-inspired models should include the emission from the space-like
parts. An example of such approach is a recent paper by the Kiev-Nantes
group \cite{Borysova:2005ng}.


\end{document}